\documentclass[% 
reprint,
 amsmath,amssymb,amsfonts,
 aps,
pra,
]{revtex4-1}

\newcommand{\trace}{\ensuremath{\text{Tr}}}

\usepackage[english]{babel}
\usepackage[utf8x]{inputenc}
\usepackage[T1]{fontenc}

\usepackage{amsmath}
\usepackage{graphicx}
\usepackage[colorinlistoftodos]{todonotes}
\usepackage[colorlinks=true, allcolors=blue]{hyperref}
\usepackage{graphicx,amssymb,amsmath,amsthm,mathtools,mathrsfs,hyperref}
\usepackage{graphicx}% Include figure files
\usepackage{dcolumn}% Align table columns on decimal point
\usepackage{bm}% bold math
\begin{document}

\preprint{APS/123-QED}

\title{Quantum control attack on quantum key distribution systems}% Force line breaks with \\

\author{Anton Kozubov}
  \email{avkozubov@corp.ifmo.ru}%Lines break automatically or can be forced with \\
\author{Andrei Gaidash}%
\author{George Miroshnichenko}
\affiliation{%
 Department of Photonics and Optical Information Technology, ITMO University,199034 Kadetskaya Line 3b, Saint Petersburg, Russia
}%
\affiliation{Quanttelecom Ltd., Saint Petersburg, 199034 Birzhevaya Line 14, Russia.}

\date{\today}% It is always \today, today,

\begin{abstract}
In this paper we present the quantum control attack on quantum key distribution systems. The cornerstone of the attack is that Eve can use unitary (polar) decomposition of her positive-operator valued measure elements, which allows her to realize the feed-forward operation (quantum control), change the states in the channel after her measurement and impose them to Bob.  Below we consider the general eavesdropping strategy and the conditions those should be satisfied to provide the attack successfully. Moreover we consider several types of the attack, each of them is based on a different type of discrimination.  We also provide the example on two non-orthogonal states and discuss different strategies in this case.

\end{abstract}

\maketitle

%\tableofcontents

Quantum key distribution (QKD) systems \cite{scarani2009security,diamanti2016practical} in the last decades extends beyond the laboratory research and go straight ahead to the market. Nevertheless a lot of vulnerabilities for real QKD systems are not covered in the various researches in both theoretical and experimental areas. Its is well-known that all implemented equipment is non-ideal. Due to this fact some attacks can be performed on real QKD systems. However the common assumption on which most existing theoretical security proofs are based is that the eavesdropper has no access to receiving and transmitting equipment. However in real life implementations she can influence the hardware of legitimate user in some way and one need to take this into consideration.

Thus possible attacks can generally be separated in two categories. The first one is the attacks on quantum states in the channel. In this field there are a huge amount of works \cite{koashi2004unconditional,renner2005simple,dupuis2016entropy,arnon2018practical,ekert1994eavesdropping,kozubov2019finite} that cover different aspects of theoretical security. All of them are based on consideration of various properties of used states. A lot of techniques were proposed for estimating the secure key since the very fist days of QKD era. However some protocols still do not have strict security proof. Some of them still have open questions, i.e. the security proof for coherent-one-way (COW) and deferential-phase-shift (DPS) protocols against  coherent attacks still do not exist.   %Among them stand subcarrier wave (SCW) QKD systems \cite{merolla1999single,merolla2002integrated,mora2012simultaneous,guerreau2005quantum,gleuim2017sideband,gleim2016secure,Miroshnichenko18, gaidash2019methods}, whose the most valuable feature is exceptionally efficient use of quantum channel bandwidth and capability of signal multiplexing by adding independent sets of quantum subcarriers around the same carrier wave \cite{mora2012simultaneous}. It makes SCW QKD systems perfect candidates as a backbone of multiuser quantum networks.

Attacks on the hardware realization or so-called "quantum hacking" belongs to the second category. 
There are huge amount of possible attacks on the hardware realization of QKD systems. One of the most crucial attack, which can be implemented to the most types of existing systems, is detector blinding  or faked-state attack \cite{lydersen2010hacking,makarov2009controlling,makarov2005faked,vakhitov2001large,lydersen2011controlling,gerhardt2011full}. A lot of experimental demonstrations of such attack was proposed for different QKD systems. Many researchers are trying to find the appropriate hardware solution. Most of them are based on some modifications of single-photon detectors (SPD). However there are still no approved experimental solutions allowing to overcome this attack. The only reliable way to deal with such Eve's strategy yet is measurement-device-independent (MDI) \cite{lo2012measurement,braunstein2012side,ma2012alternative,tamaki2012phase,liu2013experimental,yin2016measurement} or twin-field (TF) \cite{lucamarini2018overcoming} QKD systems.

In this letter we propose the approach of how to take into theoretical consideration the possibility of eavesdropper's interaction with Bob's detection node by considering quantum control attack assuming Eve may "rule" Bob's detector using appropriate states. Our approach is appropriate for protocols based on any particular kind of used states. Such clarification is important since we can construct any appropriate positive-operator valued measure (POVM). Different Eve's strategies based on various POVM constructions will be considered in this paper.

\textit{General description of the attack.}\label{General description of the attack} -- Proposed in this letter quantum controll attack relates to the class of intercept-resend attacks. However, similar attacks was considered earlier for different particular cases, for instance in the paper~\cite{lucamarini2009robust} it was considered for the case of single photons and in~\cite{gaidash2018overcoming} for the case of linearly independent states. In this letter we present generalized approach for any particular kind of states, however the most crucial case is when linearly independent states are used in the protocol. The protocol of the attack can be described as follows:

(i) Eve exchanges the channel with lossless channel and make her initial states (ancillas) $|\Psi\rangle_\mathcal{E}$ interact with the states prepared by Alice $\{|u_1\rangle_\mathcal{A},..., |u_n\rangle_\mathcal{A}\}$ using unitary operator $U_{\mathcal{AE}}$

(ii) Eve constructs the appropriate POVM and measure her states that were interacted with Alice's ones $\{|\Psi^{u_1}\rangle_\mathcal{E},\dots ,|\Psi^{u_n}\rangle_\mathcal{E}\}$

(iii) According to the measurement result she alter the states in the channel using appropriate unitary operator from polar decomposition of Kraus operator and impose them to Bob by sending them directly to him. All distinguished by Eve states should be altered and resend directly to Bob providing detection event with probability maximally close to unity. Otherwise the appropriate unitary operator (the detailed description could be found further) should be applied to the states in the quantum channel in order to prevent detection event on Bob's detector.

It can be easily noted that this attack is quite similar to the detector blinding attack.
Let us clarify the statement about detector blinding. The main idea of detector blinding attack is that the eavesdropper blinds Bob’s avalanche detectors by irradiating the receiving equipment with continuous wave (CW) laser radiation, thus switching them into a linear classical mode (or appropriate operation to control SNSPD detector). In case when Eve distinguishes the state she performs detection event at the Bob's detector. Otherwise there will be no detection event the detector.

Lets consider several possible strategies of the attack based on implementing different types of POVM to discriminate the states. In the first case of linearly independent states Eve can simply use unambiguous state discrimination measurement and identify states without errors, but with inconclusive results. The implementation of the attack and countermeasures was proposed in \cite{kozubov2019finite, gaidash2019methods}.

 Nevertheless, it is not always possible to construct errorless USD POVM. For example, there can be set of more then two states, which are linearly dependent and Eve cannot construct USD POVM.  Thus she can discriminate those states with some error probability and without inconclusive results, e.g. \cite{bennett1992experimental}.
 
 However in general case Eve may construct arbitrary POVM which contains both errors and inconclusive results (in particular she use the same detection scheme as Bob does only assuming that her hardware is perfect or at least better). Moreover this strategy is quite suitable for experimental verification.

As it was mentioned above the proposed attack is based on the fact that despite the structure of the states Eve can provide some distinguishing measurement. Also very important feature here is that we do assume the fact that efficiency of our detector is not unit and there are always losses in the channel and at the Bob's side.

Thus the attack can be described as the extension of (or general) intercept-resend attack assuming Eve can control the Bob's detector (equal to detector blinding). Since the main idea of both attacks (quantum control attack and detector blinding) is the imposing of discriminated by Eve states to Bob, they can be considered almost similar. Lets consider it as follows.

Let $\mathcal{A}, \mathcal{B}, \mathcal{E}$ be Hilbert spaces of Alice, Bob and Eve respectively and $\{|u_1\rangle_\mathcal{A},..., |u_n\rangle_\mathcal{A}\}$ is the space of states prepared by Alice.  At the beginning Alice's states and Eve's initial states (ancillas) $|\Psi\rangle_\mathcal{E}$ do not interact with each other. To obtain any information about Alice's states they should somehow interact with Eve's ancillas. It can be provided by using some unitary operation $U_{\mathcal{AE}}$ as follows:
\begin{eqnarray}
\begin{cases}
|u_1\rangle_\mathcal{A}\otimes|\Psi\rangle_\mathcal{E}\\
\vdots\\
|u_n\rangle_\mathcal{A}\otimes|\Psi\rangle_\mathcal{E}
\end{cases} \xrightarrow{U_{\mathcal{AE}}}
\begin{cases}
|\tilde{u}_1\rangle_\mathcal{B}\otimes|\Psi^{u_1}\rangle_\mathcal{E}\\
\vdots\\
|\tilde{u}_n\rangle_\mathcal{B}\otimes|\Psi^{u_n}\rangle_\mathcal{E}
\end{cases}, 
\end{eqnarray}

where $\{|\tilde{u}_1\rangle_\mathcal{B},...,|\tilde{u}_n\rangle_\mathcal{B}\}$ and $\{|\Psi^{u_1}\rangle_\mathcal{E},|\Psi^{u_n}\rangle_\mathcal{E}\}$ are the spaces of altered states in quantum channel and Eve's ancillas accordingly after Eve implies the unitary operation $U_\mathcal{AE}$. To impose any states Eve has to somehow discriminate them before it (with inconclusive results, errors, or both of them). According to the fact that this is intercept-resend attack Eve has to distinguish every single qubit. Since this Eve needs to construct the POVM with the next family of positive semi-definite operators $\{A^x_\mathcal{E}\}_{x\in\mathcal{X}}$:
\begin{gather}
 \mathbb{I}=\sum_x A^x_\mathcal{E}.
\end{gather}

Since we consider that the attack is equivalent to the detector blinding that means that Eve wants to control Bob's detector by altering the states in the channel. To do so she must resend all the states where she got detection event and be sure that Bob will also have detection event with probability equal to unity. All inconclusive results should be blocked. This can be done using unitary decomposition technique (also called polar decomposition or quantum control) \cite{shapiro1987theory,shapiro1986semiclassical,feng2018gradient,mendes2009universal} in order to alter the states in the channel according to the measurement result. Thus Eve can apply unitary decomposition to the elements of POVM as follows:
\begin{gather}
 K_\mathcal{EB}^{x}= V^x_\mathcal{BE}\left(\sqrt{A^x_\mathcal{E}}\otimes\mathbb{I}_\mathcal{B}\right),
\end{gather}
where $K_\mathcal{EB}^{x}$ is the Kraus operator, $V^x_\mathcal{BE}$ is the unitary operator, which allows to alter the states after measurement and index $x$ denotes the set of possible outcomes of implemented POVM. Such decomposition allows to realize co-called feed-forward operation (quantum control). It is quite similar to preparing the new states which form dependence on the measurement results. 
Let us consider the conditional probabilities of all possible outcomes according to the generalized measurement:
\begin{gather}
    \mathcal{P}(?|u_i)=\trace_\mathcal{E}\left(A^?_\mathcal{E}|\Psi^{u_i}\rangle_\mathcal{E}\langle\Psi^{u_i}|\right),\\
      \mathcal{P}(u_j|u_i)=\trace_\mathcal{E}\left(A^{u_j}_\mathcal{E}|\Psi^{u_i}\rangle_\mathcal{E}\langle\Psi^{u_i}|\right),
\end{gather}
here $\mathcal{P}(?|u_i)$ denotes the conditional probabilities of inconclusive results measuring $u_i$, here and further when $i=j$ we assume the case correct distinguishing, otherwise with errors, thus $\mathcal{P}(u_j|v_i)$ denotes the conditional probabilities of unambiguous discrimination if $i=j$, otherwise the conditional probabilities of errors accordingly.

  The most important thing about the discrimination for Eve is the number of her detection events. Generally it should be lower-bounded with Bob's detection rate in order that Eve can maintain the latter:
\begin{gather}\label{rate}
 \mathcal{G}_\mathcal{E} \ge   \mathcal{G}_\mathcal{B},
\end{gather}
with 
\begin{equation}
  \mathcal{G}_\mathcal{E}=N\cdot\left(\mathcal{P}(u_i|u_i)+\mathcal{P}(u_j|u_i)\right),
\end{equation}
where $N$ is the number of sent by Alice qubits.
The states (in the channel and at the Eve's side) after Eve's measurement can be presented as:
\begin{gather}
    |\tilde{u}_i\rangle_\mathcal{B}|\Psi^{u_i?}\rangle_\mathcal{E}=\frac{\sqrt{A^{?}_\mathcal{E}}|\tilde{u_i}\rangle_\mathcal{B}|\Psi^{u_i}\rangle_\mathcal{E}}{\sqrt{\mathcal{P}(?|u_i)}},\\
    |\tilde{u}_i\rangle_\mathcal{B}|\Psi^{\tilde{u}_j}\rangle_\mathcal{E}=\frac{\sqrt{A^{u_j}_\mathcal{E}}|\tilde{u}_i\rangle_\mathcal{B}|\Psi^{u_i}\rangle_\mathcal{E}}{ \sqrt{\mathcal{P}(u_j|u_i)}},
\end{gather}
where $\{|\Psi^{u_i?}\rangle_\mathcal{E},|\Psi^{\tilde{u}_j}\rangle_\mathcal{E}\}$ are the states of Eve's ancillas after her measurement. However she do not obtain any information from $|\Psi^{u_i?}\rangle_\mathcal{E}$ since all of them relate to the inconclusive results.

  When the measurement is provided Eve can implement feed-forward unitary operation as follows:
\begin{eqnarray}
 \begin{cases}
|\tilde{u}_i\rangle_\mathcal{B}\otimes|\Psi^{u_i?}\rangle_\mathcal{E} \xrightarrow{V^\emptyset_\mathcal{BE}} |\tilde{\tilde{u}}_i^?\rangle_\mathcal{B}\otimes|\tilde{\Psi}^{u_i?}\rangle_\mathcal{E} \\
|\tilde{u}_i\rangle_\mathcal{B}\otimes|\Psi^{u_j}\rangle_\mathcal{E} \xrightarrow{V_\mathcal{BE}} |\tilde{\tilde{u}}_i\rangle_\mathcal{B}\otimes|\tilde{\Psi}^{u_j}\rangle_\mathcal{E}
\end{cases}
\end{eqnarray}
here $\{ |\tilde{\tilde{u}}_i\rangle_\mathcal{B}, |\tilde{\tilde{u}}_i^?\rangle_\mathcal{B}\}$ and $\{ |\tilde{\Psi}^{u_i?}\rangle_\mathcal{E}, |\tilde{\Psi}^{u_j}\rangle_\mathcal{E}\}$ are altered states after applying unitary operators $V_\mathcal{BE}$ and $V^\emptyset_\mathcal{BE}$ in quantum channel and at Eve's side respectively, where operators $V_\mathcal{BE}$ and $V^\emptyset_\mathcal{BE}$ denotes unitary operators for conclusive result of the measurement and inconclusive outcome respectively. In case of $|\Psi^{u_i?}\rangle_\mathcal{E}$ Eve apply unitary operator $V^\emptyset_\mathcal{BE}$ to  $|\tilde{u}_i\rangle_{\mathcal{B}}$ changing it with vacuum states, otherwise she increases the absolute value of their amplitudes. It should be mentioned, $V_\mathcal{BE}$ can be chosen in a way that from the Bob's point of view there will be no additional errors introduced by Eve despite she had errors in her measurement results.
Thereby Bob cannot reveal Eve by monitoring the error rate. However, in real-world implementation there is always errors due to the non-ideality of equipment, POVM construction etc. Thus Eve should maintain both errors and detection rate. It can be done by constructing operators $V^\emptyset_\mathcal{BE}$ and $V_\mathcal{BE}$ in the way to maintain error and detection rates at the Bob's side:
\begin{gather}\label{conditions}
    \begin{cases}
    E_\mathcal{E}(\xi,\zeta)=E_\mathcal{B},\\
   \mathcal{G}_\mathcal{E}(\xi,\zeta)=\mathcal{G}_\mathcal{B},
    \end{cases}
\end{gather}
where $E_\mathcal{B}$ and $E_\mathcal{E}(\xi,\zeta)$ is the expected error rate at the Bob's side and the error rate which should be provided by Eve respectively. Here $\xi$ and $\zeta$ are the parameters to satisfy the conditions above (e.g. intensity and coding parameters changes respectively).
According to the overlapping preservation for unitary operation we assume that 
 \begin{eqnarray}
 _\mathcal{B}\langle \tilde{u}_i|\tilde{u}_j\rangle_{\mathcal{BE}} \langle \Psi^{i}|\Psi^{j}\rangle_{\mathcal{E}}= _\mathcal{B}\langle \tilde{\tilde{u}}_k|\tilde{\tilde{u}}_l\rangle_{\mathcal{BE}}\langle \tilde{\Psi}^{k}|\tilde{\Psi}^{l}\rangle_{\mathcal{E}},
 \end{eqnarray}
 where indexes $i,j$ denotes possible initial states at the Alice and Eve's side, and indexes $k,l$ denotes possible outcomes of the measurement including inconclusive result, correct distinguishing and error.  
Using proposed technique Eve can maintain both detection and error rates regarding to the presence of errors and inconclusive results at the Bob side (they always appear due to non-orthogonality of the states, losses in the channel and non-ideal equipment). Generally also different possible strategies can be applied to altering of the states. Since in this work we are interested in imposing of states distinguished by Eve. Here Eve provides the case when she exchanges all the states when she has inconclusive results with vacuum states (using appropriate unitary operator) and otherwise alters the states in the way to provide detection event with probability maximally close to unity. Thus this attack is closely related to the detector blinding (faked-state) attack, since Eve impose her discriminated states to Bob directly. However, Eve also can construct her unitary operation in that way to maintain the states in quantum channel and pretend that she does not interact with them.

\textit{Example.}\label{example} --
In this section let us consider the simplest case with two linearly independent states. The space of states prepared by Alice can be denoted as $\{|u\rangle_\mathcal{A},|v\rangle_\mathcal{A}\}$. The initial state of Eve's states (ancillas) $|\Psi\rangle_\mathcal{E}$ do not interact with states prepared by Alice at the beginning. To obtain any information about Alice's states Eve should somehow make them interact with her ancillas. It can be provided by using some unitary operation $U_{\mathcal{AE}}$ as follows:

\begin{eqnarray}
\begin{cases}
|u\rangle_\mathcal{A}\otimes|\Psi\rangle_\mathcal{E}\\
|v\rangle_\mathcal{A}\otimes|\Psi\rangle_\mathcal{E}
\end{cases} \xrightarrow{U_{\mathcal{AE}}}
\begin{cases}
|\tilde{u}\rangle_\mathcal{B}\otimes|\Psi^u\rangle_\mathcal{E}\\
|\tilde{v}\rangle_\mathcal{B}\otimes|\Psi^v\rangle_\mathcal{E}
\end{cases},
\end{eqnarray}
here $\{|\tilde{u}\rangle_\mathcal{B},|\tilde{v}\rangle_\mathcal{B}\}$ and $\{|\Psi^u\rangle_\mathcal{E},|\Psi^v\rangle_\mathcal{E}\}$ are the spaces of altered states in quantum channel and Eve's ancillas accordingly after Eve implies the unitary operation $U_\mathcal{AE}$. Lets also assume that the overlapping of Eve's states after applying unitary operation is denoted as follows:
\begin{equation}
    w=_\mathcal{E}\langle \Psi^u|\Psi^v\rangle_\mathcal{E}.
\end{equation}
The elements of POVM, since we assume that discrimination probabilities should be the same for all states, can be constructed using appropriate non-normalized vectors on the same plane as the signal states, that can be denoted as follows:
\begin{gather}
    |\phi_u\rangle_\mathcal{E}=|\Psi^u\rangle_\mathcal{E}-\mu|\Psi^v\rangle_\mathcal{E},\\
    |\phi_v\rangle_\mathcal{E}=\mu|\Psi^u\rangle_\mathcal{E}-|\Psi^v\rangle_\mathcal{E}.
\end{gather}
Then positive semi-definite operators operators of POVM can be denoted as follows:
\begin{gather}
    A_\mathcal{E}^u=\frac{1}{\delta}|\phi_u\rangle_\mathcal{E}\langle\phi_u|,\\
      A_\mathcal{E}^v=\frac{1}{\delta}|\phi_u\rangle_\mathcal{E}\langle\phi_u|,\\
      A_\mathcal{E}^?=1-A_u-A_v,
\end{gather}
here parameter $\mu$ denotes the measurement regime of POVM and $\delta$ optimizes POVM performance, reducing the probability of inconclusive result, but maintaining the positive definiteness of operator $A_\mathcal{E}^?$.
Let us consider the probabilities of all possible outcomes for one of the states according to the Born rule:
\begin{gather}
    \mathcal{P}(u|u)=\frac{1}{\delta}|_\mathcal{E}\langle\Psi^u|\phi_u\rangle_\mathcal{E}|^2=\frac{1}{\delta}(1-\mu\cdot w)^2\\
    \mathcal{P}(v|u)=|\frac{1}{\delta}|_\mathcal{E}\langle\Psi^u|\phi_v\rangle_\mathcal{E}|^2=\frac{1}{\delta}(w-\mu)^2\\
     \mathcal{P}(?|u)=1-\mathcal{P}(u|u)-\mathcal{P}(v|u),
\end{gather}
here $\mathcal{P}(?|u)$ denotes the probability of inconclusive result, $\mathcal{P}(u|u)$ and denotes the probability of unambiguous discrimination and $\mathcal{P}(v|u)$ is the probability of errors accordingly.
The vectors are chosen so that the matching conditions are met:
\begin{gather}
    \mathcal{P}(u|u)=\mathcal{P}(v|v),\\
    \mathcal{P}(v|u)=\mathcal{P}(u|v),\\
    \mathcal{P}(?|u)=\mathcal{P}(?|v).
\end{gather}

 Eve's detection rate can be obtained applying to the states the next operator $A_\mathcal{E}^D$:
\begin{gather}
    A_\mathcal{E}^D=A_\mathcal{E}^u+A_\mathcal{E}^v.
\end{gather}
Lets find the maximum eigenvalue $\lambda_{\max}$ of this operator. Optimization parameter $\delta$ is chosen then according to the next condition:
\begin{gather}
    \lambda_{\max}=1.
\end{gather}
This condition allows us to reduce the average probability of an inconclusive result and maintain the $A_\mathcal{E}^?$ operator’s positive definiteness. Denote the eigenvector of the operator as follows:
\begin{gather}
    A_\mathcal{E}^D|\chi\rangle=\lambda|\chi\rangle,\\
    |\chi\rangle=\alpha|\Psi^u\rangle_\mathcal{E}+\beta|\Psi^v\mathcal{E}
\end{gather}
In the matrix form, the equation can be presented as:
\begin{gather}
    \frac{1}{\delta}\begin{pmatrix}
    1+\mu^2-2\mu\cdot w & (1+\mu^2)w-2\mu \\
   (1+\mu^2)w-2\mu &  1+\mu^2-2\mu\cdot w
\end{pmatrix} \begin{pmatrix}
    \alpha\\
   \beta
\end{pmatrix}=\lambda \begin{pmatrix}
    \alpha\\
   \beta
\end{pmatrix}
\end{gather}
Assuming the following condition 
\begin{equation}
    \frac{2\mu}{1+\mu^2}\ge w
\end{equation}
we obtain the solution:
\begin{gather}
    \lambda_{\max}=\frac{1}{\delta}(1-w)(1+\mu)^2.
\end{gather}
Thus the probabilities of measurement outcomes can be redefined as follows:
\begin{gather}
    \mathcal{P}(u|u)=\frac{(1-\mu\cdot w)^2}{(1-w)(1+\mu)^2},\\
       \mathcal{P}(v|u)=\frac{(w-\mu)^2}{(1-w)(1+\mu)^2},\\
          \mathcal{P}(?|u)=\frac{(1+w)(1+\mu)^2}{(1-w)(1+\mu)^2}\left(\frac{2\mu}{1+\mu^2}-w\right).
\end{gather}
Lets consider different POVM strategies. The first case is errorless USD POVM can be obtained if $\mu=w$. Then the probabilities of different outcomes has the next form:
\begin{gather}
  \mathcal{P}(u|u)= 1-w,\\
  \mathcal{P}(v|u)=0,\\
  \mathcal{P}(?|u)=w
\end{gather}
However, if Eve implement such discrimination without errors, she can be revealed by increasing the number of states, since the discrimination probability is a function of number of states.

Another degenerate case is when her discrimination POVM does not produce any inconclusive results (measurement in Breidbard basis\cite{bennett1992experimental}) can be obtained if the following condition is satisfied
\begin{gather}
       \frac{2\mu}{1+\mu^2}=w,\\
       \mu=\frac{1-\sqrt{1-w^2}}{w}.
\end{gather}
In this case the detection rate $\mathcal{G}_\mathcal{E}$ is equal to unity, so she can satisfy the condition in eq.~\ref{rate}. Then the probabilities of different can be calculated using the following expressions: 
\begin{gather}
    \mathcal{P}(?|u)=0,\\
   \mathcal{P}(u|u)=\frac{w^2}{2(1-\sqrt{1-w^2})},\\
    \mathcal{P}(v|u)=\frac{w^2}{2(1+\sqrt{1-w^2})}.
\end{gather}

Nevertheless in this case Eve produces a huge amount of errors and can be easily revealed by error rates.

Thus the only way for her to provide the successful attack is to implement the general case, then her POVM has both errors and inconclusive results. It can be done by changing a parameter $\mu$. The simplest way to implement such POVM is just to construct the same as Bob's one. If she does so the condition in eq.~\ref{rate} will be naturally satisfied, since Eve has ideal (or at least better) equipment. To provide the attack successfully she should satisfy the next condition:
\begin{gather}
    N\frac{(1-\mu\cdot w)^2+(w-\mu)^2}{(1-w)(1+\mu)^2}\ge\mathcal{G}_\mathcal{B},
\end{gather}
It is quite obvious that $\mu$ can be chosen in that way to satisfy the condition. The states after Eve measurement can be presented as:
\begin{gather}
    |\tilde{u}\rangle_\mathcal{B}|\Psi^{u?}\rangle_\mathcal{E}=\frac{\sqrt{A^{?}_\mathcal{E}}|\tilde{u}\rangle_\mathcal{B}|\Psi^{u}\rangle_\mathcal{E}}{\sqrt{\mathcal{P}(?|u)}},\\
    |\tilde{u}\rangle_\mathcal{B}|\Psi^{\tilde{u}}\rangle_\mathcal{E}=\frac{\sqrt{A^u_\mathcal{E}}|\tilde{u}\rangle_\mathcal{B}|\Psi^{u}\rangle_\mathcal{E}}{ \sqrt{\mathcal{P}(u|u)}},\\
    |\tilde{u}\rangle_\mathcal{B}|\Psi^{\tilde{v}}\rangle_\mathcal{E}=\frac{\sqrt{A^v_\mathcal{E}}|\tilde{u}\rangle_\mathcal{B}|\Psi^{u}\rangle_\mathcal{E}}{ \sqrt{\mathcal{P}(v|u)}},
\end{gather}

After measuring her states Eve should directly resend all her discriminated states (both with error or without) with probability equal or close to unity. As it was shown in previous  section she should apply unitary trace-preserving operators $V_\mathcal{BE}$ and $V^\emptyset_\mathcal{BE}$ which should be constructed according to eq.~\ref{conditions} as follows:
\begin{eqnarray}
 \begin{cases}
|\tilde{u}\rangle_\mathcal{B}\otimes|\Psi^{u}\rangle_\mathcal{E}\xrightarrow{V_\mathcal{BE}}|\tilde{\tilde{u}}\rangle_\mathcal{B}\otimes|\tilde{\Psi}^{u}\rangle_\mathcal{E}\\
|\tilde{u}\rangle_\mathcal{B}\otimes|\Psi^{u?}\rangle_\mathcal{E}\xrightarrow{V^\emptyset_\mathcal{BE}}|\tilde{\tilde{u}}^?\rangle_\mathcal{B}\otimes|\tilde{\Psi}^{u?}\rangle_\mathcal{E}\\
|\tilde{u}\rangle_\mathcal{B}\otimes|\Psi^{v}\rangle_\mathcal{E}\xrightarrow{V_\mathcal{BE}}|\tilde{\tilde{u}}\rangle_\mathcal{B}\otimes|\tilde{\Psi}^{v}\rangle_\mathcal{E}
\end{cases},
\end{eqnarray}
here $\{|\tilde{\tilde{u}}\rangle_\mathcal{B}, |\tilde{\tilde{u}}^?\rangle_\mathcal{B}\}$ and $\{|\tilde{\Psi}^{u}\rangle_\mathcal{E},|\tilde{\Psi}^{u?}\rangle_\mathcal{E},|\tilde{\Psi}^{v}\rangle_\mathcal{E}\}$ are the states after applying operators $V_\mathcal{BE}$ and $V^\emptyset_\mathcal{BE}$ in quantum channel and Eve's ancillas accordingly to the measurement outcome as it was shown in previous section.
The operator can alter the states in any appropriate way.

The possible countermeasures against such attack are closely related with countermeasures against faked-state attack. Since Eve should maintain the overlapping and detection rate at the Bob's side she should increase the of the states. Thereby the possible way to reveal Eve is to use photon-number-resolving (PNR) detectors which allows to monitor the statistics of the radiation (e.g. the common approach can be seen in \cite{gaidash2016revealing}). Also the countermeasure from \cite{chistiakov2019controlling} can be implemented. Another possible solution is presented in \cite{ko2018advanced}, where coincidence detection events are monitored.

\textit{Conclusion.}\label{conclusion} --
In this paper we present the technique which allows Eve to provide so-called quantum control attack. The attack is based on the fact, that Eve wants to impose discriminated states to Bob and thus obtain up to 100\% of the shared key. We consider the general case then Eve can implement any appropriate POVM which allows her (or not) to obtain the necessary number detection events to maintain both detection and error rates at the Bob's side. The cornerstone of this paper is that considered attack can be implemented to any states. However not every suggested strategy can be realized successfully if the proposed conditions are not satisfied by Eve. Nevertheless, such technique allows us to take into theoretical consideration the detector blinding quantum hacking attack. Moreover we propose the possible countermeasures for different types of the attack. This approach allows us to make theoretical security proofs more accurate not only for ideal-world considerations but the real-world implementations.

This work was financially supported by Government of Russian Federation (Grant 08-08).

\bibliography{bibliography1}

\end{document}